# BIM, DIGITAL TWIN AND CYBER-PHYSICAL SYSTEMS: CROSSING AND BLURRING BOUNDARIES

Dean Douglas[1] (PhD Candidate), Dr. Graham Kelly[2] and Prof. Mohamad Kassem[1]
[1] Northumbria University, Newcastle-Upon-Tyne, UK
[2] BIM Academy, Newcastle Upon Tyne, United Kingdom

## ABSTRACT

Digital Twin in construction and the built environment have started to attract the attention of researchers and practitioners in recent times. Its anticipated value proposition is focussed on its capability of generating new understanding and insights into an asset at all stages of its lifecycle, exploiting diverse data sets from a multitude of sources and professions, in real or near real-time. However, there is still a significant debate about the delineation (i.e. communalities and differences) between digital twin and other related concepts, particularly Building Information Modelling (BIM) and Cyber-Physical Systems (CPS). To date, this debate has been confined to social media discussions, insights blogs and position papers. This paper addresses this challenge using a systematic review. The aim is to investigate communalities and differences between the three concepts, Digital Twin, BIM and CPS. The results of this paper are expected to foster the discussion around this theme within construction and the built environment.

## INTRODUCTION

To date the digital twin concept predominantly has been applied within the manufacturing and production industry. However, the potential areas of application of digital twin are wide ranging from physical assets, which span from components of machineries to large scale infrastructure assets (e.g. roads and railways), to social and economic behaviours (Wang *et al.*, 2016), medicine and pharmaceuticals (Geris *et al.*, 2018; Torkamani *et al.*, 2017), or a composite of these such as in a construction project comprised of both physical assets and social systems elements (Sacks *et al.*, 2020).

Recently, the discussion of the digital twin concept in construction and built environment started to generate intense discussions in social media around its delineation, definition and interactions with other established concept such as BIM and CPS. While a thematic coding of views on social media is outside the scope of this paper, a tentative gradation of opinions from discussions on social media categorises views into three clusters: (1) the *'unconvinced'*, those who are dismissing the digital twin concept and depicting it as either a rebranding of BIM or a marketing invention; (2) the *'undecided'*: those who are calling for an investigation of the digital twin capabilities and uses within the built environment, and how these differ from those enabled by BIM and CPS, before legitimising the concept; and (3) the *'committed'*: those who have firmly accepted the digital twin concept based on various propositions with the main one being the sensorial connections between the physical assets and their digital replicas. Motivated by this debate and the lack of sufficient research-based insights about the digital twin concept and its communalities and differences with potentially overlapping or complementary concepts such as BIM and CPS, we proposed to investigate this gap in knowledge. This understanding of potential relationships between digital twin, BIM and CPM concepts and technologies is important to the field researchers who are looking to address built environment challenges through the application of digital twin. While the paper discusses the relationship between the digital twin and both BIM and CPS, it outlines the prevalent arguments and assumptions made in the relevant peer reviewed studies.

## SYTEMATIC REVIEW

This research was informed by the literature selected through the search process depicted in Figure 1 using search terms such as: *"Digital Twin" AND (infrastructure OR "asset management" OR "facilities management" OR "built environment" OR construction OR building OR handover OR maintain OR maintenance OR "building lifecycle" OR "asset lifecycle" OR "project lifecycle" OR architecture OR engineering OR "smart cities" OR "urban planning" OR asset OR design OR operation).*

The databases searched included the Web of Science, Scopus and Science Direct. The following exclusion criteria for the literature search were applied at the screen and eligibility stages.

**Screening**

- Repeated Article
- Not written in English
- Predates the conception of digital twin; (Glaessgen and Stargel, 2012) - this is widely considered to be the first application of digital twin.

**Eligibility**

- They offer no definition of digital twin.
- Did not include at least one of the following: applications, elements, supporting technologies or adoption.
- Digital twin is only discussed in the conclusion as a further step, or as a potential application/addition to the paper topic.
- References to digital that were considered to be speculative and not grounded.

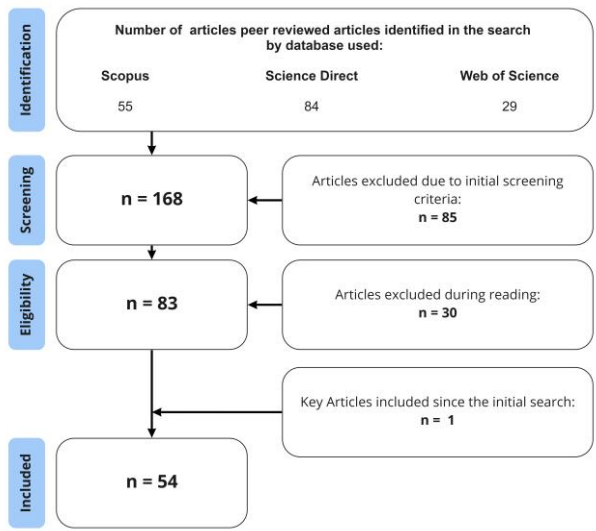

*Figure 1 The literature search process*

## DIGITAL TWIN AND BIM

An issue that has become apparent is the confusion about the relationship between digital twin and BIM. There are three key understandings. The first portrays Digital Twin as a continuation of BIM that would incorporate new technological developments. The second argues the two concepts differ significantly in their characteristics and capabilities and they should be considered as distinct concepts. In between these two stances, the third understanding consider the technologies as complementary and can co-exist in their applications and uses over the lifecycle of an asset. Using a subset of the studies identified in our review, we critically discuss each of these understandings:

**Digital Twin as a continuation of BIM**

Digital Twin is seen as a continuation of BIM in (Boje *et al.*, 2020), where BIM is considered as the potential starting platform to initiate an evolutionary three-tier approach to the development of digital twin in the built environment. The three levels, depicted in Figure 2 include:

**Generation #1 - monitoring platforms**
Monitoring platforms enable the sensing of an asset utilising reporting and limited analysis capabilities. Legacy digital models are used in combination with sensors is the first step towards real-time integration of site sensing and digital models.

**Generation #2 - intelligent semantic platforms**
Creation of enhanced monitoring platform can be enabled using a common web language framework to integrate digital twin with all its IoT devices. The digital twin has limited intelligence relying on embedded knowledge rules and separate AI-enabled simulations and predictions. This generation entails significant human interactions to enact optimisation and operation of the digital twin.

**Generation #3 - agent-driven socio-technical platforms.**
A fully semantic digital twin that is capable of leveraging acquired knowledge with the use of AI-enabled agents. A variety of digital technologies and techniques such as machine learning, deep learning, data mining and data analysis are employed to create a digital twin that is self-reliant, self-updatable and self-learning. Operation becomes fully autonomous requiring only human supervision.

Current implementation of BIM is conceived by the authors in this model to be at the point of initiation of Generation #1. Boje *et al.*, (2020) support this assumption by arguing that BIM in its current state cannot deliver the information requirements throughout an asset lifecycle and that even with extensions to its current capabilities it is unlikely to be able to perform more complex computations such as prediction and optimisation.

In the same vein, Akbarieh *et al.*, (2020) state that BIM provides a static representation of the material of an asset, whereas a Digital Twin represents information that is an accumulated over the course of an assets lifecycle including information about condition and maintenance. They also argue BIM can be used to create Digital Twin alongside additional tools.

However, these two stances about the difference between BIM and Digital Twin are questionable as the understanding depends on the assumed definition for BIM. For example, it could be argued that Digital Twin Generation #1 (monitoring platforms) can be considered either as a capability that can be fulfilled by BIM if BIM is referred to as both a technology and an information management methodology, or as a complementary capability of BIM if BIM is conceived as requiring 3D model whereas monitoring and analysis of physical assets through sensing does not necessarily require a 3D model. In either cases, these two stances claiming that digital twin is a continuation of BIM can be challenged.

In this category of thought that consider BIM as a starting platform for digital twin Lu, Xie, *et al.*, (2020) take the view that building information models of assets, utilising the Industry Foundation Classes IFC) schema, could be the basis for an operation and maintenance digital twin, that is capable of integrating numerous disparate data sources. However, for this to be realised, the authors argue that an operation and maintenance extension to the current IFC would be required. In addition to the use of IFC (Lu, Xie, *et al.*, 2020) also suggest that Construction Operations Building Information Exchange (COBie) could aid in the formatting of data to be integrated into the digital twin, given its wide adoption and ability to adapt to suit applications. However, there are limitations to this approach due to the challenges of monitoring every single asset within a system. Hence, its application would be limited to critical assets where relevant data is likely to be available. Nevertheless, these views of BIM and digital twin seems to limit BIM to its IFC schema or COBie that would serve as repository to host the sensors' data of a digital twin which is also can be considered as reductive of the BIM concept. For example, numerous BIM uses or model uses have integrated sensors data into a BIM model (as in Bamakan *et al.*, 2020; Quinn *et al.*, 2020; Riaz *et al.*, 2014) which clearly challenge this stance about digital twin being seen as a continuation of BIM.

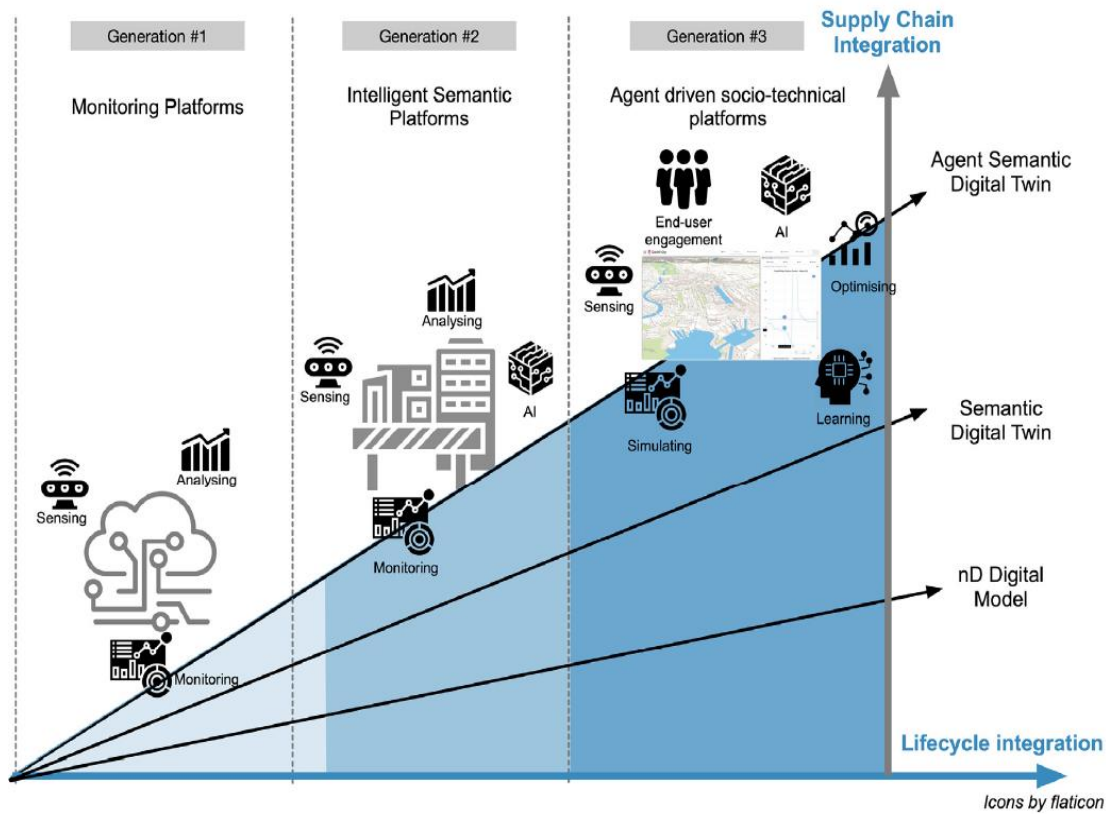

*Figure 2 The three-tier generation evolution of the construction digital twin (Boje et al., 2020)*

*Table 1 A detailed comparison of BIM and digital twin of building (Khajavi et al, (2019)*

| Concept / Differentiator | Application Focus | Users | Supporting technology | Stage of life cycle |
|---|---|---|---|---|
| BIM | Design visualisation and consistency, clash detection, lean construction, time and cost estimation, stakeholders interoperability | AEC, Facility Manager | detailed 3D model, common data environment, industry foundation class, COBie | Design, Construction, use (maintenance), demolition |
| Digital Twin | Predictive maintenance, tenant comfort enhancement, resource consumption efficiency, what-if analysis, closed-loop design | Architect, Facility Manager | 3D model, wireless sensor network, data analytics, machine learning | use (operation) |

**BIM & Digital Twin as two distinct concepts**

Another understanding, represented by a growing amount of studies, argues that there are several succinct differences between digital twin and BIM that are deemed to be significant to classify them as two separate concepts. Sacks et al (2020) suggests that a BIM model that is created during the design and construction of an asset may provide an accurate as built model; however, it still deficient be considered a digital twin which is conceived as a representation of an asset that is continually updatingwith the state of the asset. Sacks et al (2020) further elaborate that even BIM models that are used in the operation and maintenance of an asset are updated reactively and are not intended to create the short term feedback loop available in digital twin. This stance is basically raising the promptness of feedback about the state of an asset as a key determinant to distinguish a BIM from a digital twin. In the same vein, Lee et al. (2012) define Digital Twin as virtual replica of physical assets, process, system or service which represents the properties (e.g., the geometry of assets), the condition (e.g., resource status), and the project's performance. However, the authors define BIM – citing US National BIM Standard – as a digital representation of what will be built, and argues citing Tchana et al. (2019) that the core technology of the Digital Twin is data synchronization between the physical asset and its virtual representation. Hence, this understanding is limiting BIM to a virtual replica of the built asset while there are many studies that already investigate and demonstrate the ability of building information models to integrate real-time data from sensor networks. (Alves *et al.*, 2017; Chen *et al.*, 2014; Riaz *et al.*, 2017).

Theiler *et al* (2020) suggests that while the concept of Digital Twin is highly associated with the BIM approach in the way they share a number of similarities surrounding the metaization of asset information, Digital Twin not only covers metamodeling of an asset but also the integration of simulation capabilities, which would then require the prototyping of asset systems in a simulation software.

Khajavi *et al*, (2019) explores the differences between digital twin and BIM further, comparing them in attributes such as: focus of applications, the users, the supporting technologies, and the most common stage of life cycle (Table 1). As to the focus of application the authors argue that BIM application are mainly confined in the design and construction phases of an asset's life cycle to facilitate communication between stakeholders, reduce design error, and monitor time and cost of construction. Whereas, a digital twin can be applied to conduct what-if analysis, to enhance occupant comfort and to transfer lessons learned from one asset into the future design of another, referred to as closed-loop design. It can also be seen in Table 1, that facility managers have been said to use both BIM and digital twin in the in-use phase of an asset's life cycle, however, the processes that they inform are quite different. Where BIM is used to better inform maintenance; digital twin is used to enhance an assets operation. Also this distinction between BIM and digital twin is not defensible as there is a growing study domain, often referred to as 'BIM for facilities management', where BIM is used for both asset maintenance and operation.

**BIM & Digital Twin as complementary concepts**

This third understanding argue that BIM and digital twin are complementary concepts and can be used in tandem at any stage of an asset lifecycle. Pan and Zhang (2021) draws attention to the inclusion of BIM, Internet of Things (IoT) and data mining techniques into the digital twin as a means of delivering smarter construction services. In the example given by Pan and Zhang (2021), multiple building information models of an asset at different stages of its lifecycle (i.e. as design and as built) are utilised to enrich the digital twin. Hou *et al* (2021) takes the stance that during the pre-construction phase of an asset, where it is not possible to have a Digital Twin -due to the lack a physical counterpart- that a BIM model created at this point along with other information and communication technology used can become the information basis for a Digital Twin. Going further to say that during the construction phase that a 4D BIM model coupled with real time sensor data of site activity can be generated to simulate the construction environment. Errandonea *et al* (2020) takes the view that BIM coupled with the finite element method could be utilised in a Digital Twin as a means of bridging the gaps in available data. Another example by Abdelmegid *et al* (2020) suggested to investigate the integration of BIM and digital twin for simulation modelling and lean-based planning and control applications given the overlaps between phases of implementation and ease of integration with management practices as planning and control. Similarly Shirowzhan *et al* (2020) highlights the challenges faced by Digital Twin development stressing the importance of integration and interoperability of BIM along with other prominent technologies and processes such as Geographic Information System (GIS), Virtual/Augmented Reality (AR/VR) and the Internet of Things (IoT) as this will better enable the modelling of interacting subsystems.

## DIGITAL TWIN AND CYBER PHYSICAL SYSTEMS: BLURRING DELINEATION

Another area of ambiguity that is causing debate is the potential confusion between the concepts of digital twin and cyber-physical systems, their relationship, terminology and origins. In many instances the terms were used near interchangeably to relate to the same concept, while in other instances they have been stated to be two related yet entirely separate concepts.

Numerous descriptions that define the relationship between digital twin as a component or result of the CPS process. With Hoffmann Souza *et al* (2020) offering that digital twin as a technology provided by the creation of a CPS, that can then become a means of simulating and monitoring an asset. Sun *et al* (2020) goes further to begin to understand the relationship between them stating that CPS is the core of digitisation, with several different technologies including IoT, AI, Big Data analytics and cloud computing all contributing to its creation. It is through this process of combining these technologies and the collecting of big data that a digital twin is said to be created. Wu *et al* (2020) directly attributes the confusion of definitions and relationships between CPS in manufacturing and Digital Twin to the lack of an agreed upon definition of CPS, which in turn is said to be due to the relative new-ness of the term.

Leng *et al* (2020) seeks to understand and define the relationship between CPS, digital twin and IoT. They argue that CPS encompasses both the digital twin as the digital replica of an asset and the IoT as the many sensors and controls deployed in an asset that provides the bridge between the digital and physical entities. To the same ends, Y. Lu *et al* (2020) provides Figure 3 which maps out the interrelationship between digital twin, CPS, and IoT: IoT is the architecture of sensors that harvest data from the physical asset; and the CPS encompasses the physical asset, the digital twin of that asset and the IoT that is the bridge between them. It is also seen here that the digital win is not only a single digital replica of an asset but rather an ecosystem of digital replicas.

Contrary to this, there is also the view that CPS and digital twin are not interconnected technologies that form part of a wider system but instead as two similar yet separate concepts of development. Tao *et al* (2019) investigates this topic extensively, stating that both CPS and digital twin originate from two separate origins that have influenced their structure, development and areas of implementation. Digital twin originates from a complex engineering background that sought to understand the interdependencies between engineering systems, whereas CPS was born of a more scientific background that sought to efficiently describe complex systems using traditional IT terminology (Tao *et al.*, 2019).

The two concepts do share a number of common characteristics as set out by Tao *et al,* (2019), with both consisting of a physical asset and a digital/cyber element where the physical is tasked with the harvesting of asset data to be relayed to the cyber and similarly to actuate the commands of the cyber after analysis and decision making. In order to achieve this, the cyber/digital employs a number of applications and technologies to generate insights to enhance decision making within a system. These two parts create a continual cyclical feedback loop of "physical world-digital world-physical world" (Lu *et al.*, 2019).

The hierarchical structure of both digital twin and CPS are considered to be formed of the same three levels; the unit level, system level, and system of systems level as can be seen in Figure 4 (Qi *et al.*, 2018). What is also suggested here is that the creation of a comprehensive digital twin or CPS must start in the same way, with a bottom-up approach, starting with the definition of unit-level processes and assets, slowly building until the second level is achieved and a system is modelled either in part or full.

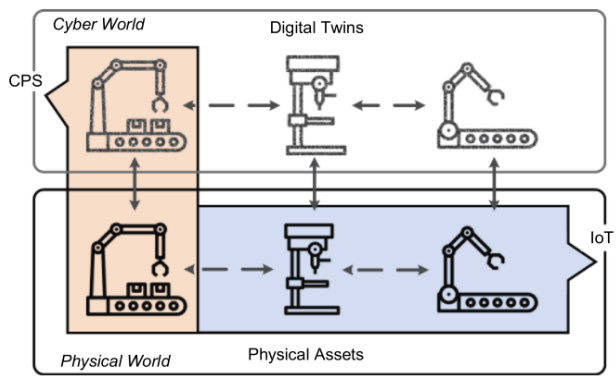

*Figure 3 The relationship between Digital Twin, CPS and IoT (Y. Lu et al., 2020*

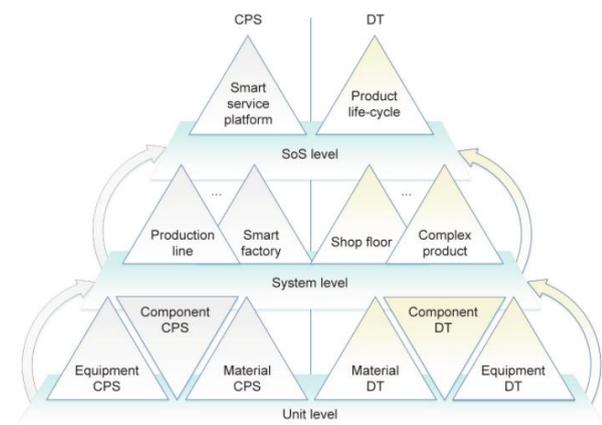

*Figure 4 Hierarchical levels of CPS and DTs in manufacturing (Qi et al., 2018)*

Finally, beginning to relate these different systems to one another as to achieve the third level of a system of systems. While taking a similar stance that Digital Twin and CPS are two separate yet highly related concepts, Ciano *et al* (2020) outlines evidence that suggests that CPS is an evolution of embedded systems where the CPS network can coordinate and integrate computation and physical processes. Computational components serve to monitor, detect and activate physical elements. They state that the focus of contemporary research regarding CPS focuses on the link between the embedded systems of the physical layer and the application layer. According to the author, it is here that CPS and Digital Twin could benefit one another, as Digital Twin could become the bridge between the physical and applications layers with the Digital Twin containing all the necessary information needed for the working a of a CPS. However, these propositions suggest that while CPS and Digital Twin can benefit from there interrelation, they are not dependant upon one another and they can both exist independently of the other.

While CPS and digital twin have a number of similarities, they are also said to have some defining differences, primarily in their emphases. Tao *et al* (2019) explores this for digital twin citing Schleich *et al* (2017) and DebRoy *et al* (2017) who suggest that the Digital Twin are not only comprised of a highly consistent model of the physical in terms of geometry and structure but also that it is capable of simulating its spatiotemporal status, its behaviours, and functions. Going on to further highlight the focus on the creation and definition of a high-fidelity virtual model with a one-to-one correspondence between physical and virtual components, with the high-fidelity of the replica enabling the digital twin to understand and predict occurrences in the physical (Tao *et al.*, 2019). Whereas with CPS Tao *et al* (2019) provides overviews of CPS from Wang *et al* (2015) which suggests that at its core that CPS strives to add new capabilities to the physical system by utilising the intensive interaction of computation and communication into the physical process. Additionally, CPS is thought to follow the principles of collaboration of computing, communication, and control; which results in a one to many correspondence and gives greater attention to the data exchange loop between the physical and cyber systems, and thus placing the network of sensors and actuators at its core (Tao *et al.*, 2019). It is because of these differences in emphasis that there can be a disparity in the aspects captured and modelled in the levels of the structural hierarchy (Figure 4), while the structure itself remains the same.

## DISCUSSION

Through the course of this paper an attempt has been made to provide research-based insights drawn from relevant peer reviewed literature, into the disparity of how digital twin is discussed in relation to the already established notion of BIM in construction and the built environment and the CPS concept.

The inconsistencies in digital twin discussions were found significant to the extent it was possible to identify three different understandings of about its delineations with BIM. The first understanding with studies such as Boje *et al* (2020) and Lu, Xie, *et al* (2020) outlines routes to digital twin realisation that are seen to be the continuation of the BIM methodology, by incorporating new technological advances. The second understanding considers BIM and digital twin as two distinct concepts based on dissimilarity in aspects as: their abilities to incorporate real time data feedback loops (Sacks *et al.*, 2020), the difference in the professionals that make use of them, their current capabilities and areas of application and the stage of an assets lifecycle (Khajavi *et al.*, 2019). Both understanding were not fully resolute considering the counter evidence from the peer reviewed literature. A third understanding suggested that BIM and digital twin are two complementary concepts that can be simultaneously used to develop new capabilities such simulation of performance and lean planning and control.

The potential delineations of digital twin and CPS seem further apart with one another. On one side it is argued that digital twin and CPS are both parts of a wider ecosystem of digital technologies and processes while on another, the two concepts are considered distinct but with a number of common characteristics.

A number of studies such as Hoffmann Souza *et al* (2020), Sun *et al* (2020), Y. Lu *et al* (2020) and Leng *et al* (2020) have taken the stance that digital twin is an integral part of the wider CPS system as it forms the virtual representation of a physical asset that utilises a plethora of technologies and applications such as IoT, AI, Big Data analytics and cloud computing.

Contrary to this, Tao *et al* (2019) suggest that digital twin and CPS are two individual concepts, yet, share a number of similarities such as; the structure of hierarchical levels, the components they are comprised of, and also that they both utilise the variety of applications and technologies to meet their intended purposes. However, the two are said to differ greatly in their development emphasis, which in turn creates disparity in the aspects captured and modelled within the respective concepts.

Understanding the relationship between digital twin and CPS could greatly alter not only the language and terminology used when discussing the components of a digital twin, but also its applications in the future.

If the built environment is to realise the potential benefits of digitalisation, future research should strive to build consensus about the communalities, differences and interactions between the digital twin, BIM and CPS concepts. If this challenge is not addressed, the risk of compromising the reusability of knowledge that is available in the industrial and academic literature and its retrieval become a real prospect. This will partly contribute to slowing down the digitalisation of the construction sector at a time when the sector is facing new challenges and is being asked to be more efficient than ever.

## CONCLUSIONS

Differences and commonalties between digital twin, BIM and CPS are becoming a recurrent topic of discussion in academic and industry settings. To date, this issue has not been addressed in a research-oriented approach. This paper addressed this gap using a systematic review. We identified three different understandings in relation to digital twin and BIM. All the three stances could be challenged with counter-evidence from the literature. While this paper does not attempt to develop a new definition of either digital twin, BIM or CPS, it was successful in highlighting the debate and the different views about the three concepts. It contributed to classify emerging views, highlighting the differences between them, and calling for further investigation into this challenge. Understanding the delineation between the three concepts is important to harness knowledge and accelerate digital transformation within the construction sector and the built environment.